\pdfoutput=1
\documentclass[submission,copyright,creativecommons]{eptcs}
\providecommand{\event}{ICE 2020}
\title{Structural Equivalences for Reversible Calculi of Communicating Systems\\ (Oral communication)}
\author{Clément Aubert
\institute{School of Computer and Cyber Sciences,\\ Augusta University, USA}
\email{\href{mailto:caubert@augusta.edu}{caubert@augusta.edu}}
\and
Ioana Cristescu
\institute{Tarides,\\ Paris, France}
\email{\href{mailto:ioana@tarides.com}{ioana@tarides.com}}
}
\def\titlerunning{Structural Equivalences for Reversible Calculi of Communicating Systems}
\def\authorrunning{C. Aubert and I. Cristescu}

\usepackage{mathtools}
\usepackage{amssymb}
\usepackage{ebproof}

\usepackage{thm-restate}
\usepackage[inline]{enumitem} %
\usepackage{stmaryrd}

\usepackage{cases}
\usepackage{empheq}

\usepackage[OMLmathsfit]{isomath}

\usepackage{multicol}

\usepackage{xspace}
\usepackage{csquotes}

\usepackage{tikz}
\usetikzlibrary{%
	arrows.meta, %
	calc,
	backgrounds,
	fit,
	patterns}
\usepackage{graphicx}

\usepackage{stackengine,scalerel,graphicx,trimclip,wasysym}

\usepackage{xcolor,
	colortbl}

\usepackage{microtype}

\savestack\zigzagtextstyle{\vphantom{()}\scalebox{.86666}[1]{\kern-.5pt\AC\kern-.5pt}}
\savestack\onelinetextstyle{\vphantom{()}\scalebox{1}[1]{\kern-1pt{$-$}\kern-1pt}}
\savestack\twolinetextstyle{\vphantom{()}\scalebox{1}[1]{\kern-1pt{$=$}\kern-1pt}}
\savestack\ZigZagtextstyle{%
 $\vphantom{()}\smash{\raisebox{-.002em}{$\vcenter{\hbox{%
 \stackengine{-.805em}{\zigzagtextstyle}{\zigzagtextstyle}{O}{c}{F}{F}{S}}}$}}$}
\newlength\repwidth
\repwidth=\wd\zigzagtextstylecontent\relax
\newcommand\rightarrowhead{\clipbox{4pt -2pt 0pt -2pt}{$\rightarrow$}}
\newcommand\Rightarrowhead{\clipbox{4pt -2pt 0pt -2pt}{$\Rightarrow$}}
\newcommand\rrightarrowhead{\clipbox{5.5pt -2pt 0pt -2pt}{$\rightarrow\kern-8.5pt\rightarrow$}}
\newcommand\RRightarrowhead{\clipbox{5.5pt -2pt 0pt -2pt}{$\Rightarrow\kern-8.5pt\Rightarrow$}}
\newcommand\fXarrowtextstyle[2]{$\vphantom{()}\smash{\vcenter{\hbox{\kern-.03\repwidth%
 \clipbox{-.03\repwidth{} 0pt .527\repwidth{} 0pt}{#1}}}}#2$}
\savestack\fzigarrowtextstyle{\fXarrowtextstyle{\zigzagtextstyle}{\rightarrowhead}}
\savestack\fZigarrowtextstyle{\fXarrowtextstyle{\ZigZagtextstyle}{\Rightarrowhead}}
\savestack\flinearrowtextstyle{\fXarrowtextstyle{\onelinetextstyle}{\rightarrowhead}}
\savestack\fLinearrowtextstyle{\fXarrowtextstyle{\twolinetextstyle}{\Rightarrowhead}}
\savestack\fzzigarrowtextstyle{\fXarrowtextstyle{\zigzagtextstyle}{\rrightarrowhead}}
\savestack\fZZigarrowtextstyle{\fXarrowtextstyle{\ZigZagtextstyle}{\RRightarrowhead}}
\savestack\fllinearrowtextstyle{\fXarrowtextstyle{\onelinetextstyle}{\rrightarrowhead}}
\savestack\fLLinearrowtextstyle{\fXarrowtextstyle{\twolinetextstyle}{\RRightarrowhead}}

\newcommand\oneline{\scalerel*{\onelinetextstyle}{()}}

\newcommand\flinearrow{\scalerel*{\flinearrowtextstyle}{()}}

\newcommand\linearrow[1][]{\Xarrow{\oneline}{\flinearrow}{#1}{-.65}}

\newcommand\Xarrow[4]{\ThisStyle{\mathrel{\Xarrowhelp{#1#2}{#3}{#1}{#4}}}}
\newcommand\Xarrowhelp[4]{%
 \setbox0=\hbox{$\SavedStyle#1$}%
 \setbox2=\hbox{$\SavedStyle_{\,\,#2\,\,}$}%
 \ifdim\wd0<\wd2\relax\Xarrowhelp{#3#1}{#2}{#3}{#4}%
 \else\stackengine{#4\LMex}{\copy0}{\copy2\,}{O}{c}{F}{T}{S}\fi%
}%
\newcommand{\efs}[1][]{\linearrow[#1]}

\newcommand{\redl}[1]{\efs[#1]}

\newcommand{\congru}{\equiv}

 \newcommand{\fork}{\curlyvee}
\newcommand{\bs}{\backslash}

\newcommand{\fn}[1]{\ensuremath{\mathrm{fn}(#1)}}

\newcommand{\out}[1]{\overline{#1}}

\newcommand{\BNFsepa}{\enspace \Arrowvert \enspace}

\newcommand{\namelist}[1]{\overrightarrow{#1}}

\AtBeginDocument{

	\def\itemautorefname~#1\null{\textcolor{darkgray}{\sffamily\bfseries\upshape\mathversion{bold}#1.}\null}
}

\newcommand*{\eg}{e.g.\@\xspace}

\makeatletter
\newcommand*{\etc}{%
 \@ifnextchar{.}%
 {etc}%
 {etc.\@\xspace}%
}
\makeatother

\renewcommand{\footnotesize}{\fontsize{8pt}{10pt}\selectfont} %

\newtheorem{theorem}{Theorem}
\newtheorem{definition}[theorem]{Definition}
\newtheorem{lemma}{Lemma}

\newcommand\hypo{\Hypo}
\newcommand\infer{\Infer}
\AtBeginDocument{
	\hypersetup{
		breaklinks=true,
		pdfborder={0 0 0},
		pdftitle={Structural Equivalences for Reversible Calculi of Communicating Systems},
		pdfauthor={Clément Aubert, Ioana Cristescu}
	}
}

\begin{document}
\maketitle

\begin{abstract}
	The formalization of process algebras usually starts with a minimal core of operators and rules for its transition system, and then relax the system to improve its usability and ease the proofs.
	In the calculus of communicating systems (CCS), the structural congruence plays this role by making \eg parallel composition commutative and associative: without it, the system would be cumbersome to use and reason about, and it can be proven that this change is innocuous in a precise technical sense.
	For the two reversible calculi extending CCS, the situation is less clear: CCS with Communication Keys (CCSK) was first defined without any structural congruence, and then was endowed with a fragment of CCS's congruence.
	Reversible CCS (RCCS) made the choice of \enquote{backing in} the structural equivalence, that became part of the \enquote{minimal core} of the system.
	In this short oral communication, we would like to re-consider the status and role of the structural congruence in general, to question its role in RCCS in particular, and to ask the more general question of the structural equivalences legitimacy.
\end{abstract}

\section{Introduction}

It is often useful to have a look at the fundamentals of your research field.
For instance, a classical textbook states that constructing a process algebra often uses a two-stages approach~\cite[p.~V]{Bergstra2001}:

\textquote{In a process-algebraic approach to system verification, one typically writes two specifications.
	One, call it SYS, captures the design of the actual system and the other, call it SPEC, describes the system's desired \enquote{high-level} behavior.
	One may then establish the correctness of SYS with respect to SPEC by showing that SYS behaves the \enquote{same as} SPEC.}

This is generally the approach taken to define the calculus of communicating systems (CCS)~\cite{Milner1980}.

\subparagraph{Stage 1: \texorpdfstring{SYS\textsubscript{CCS}}{SYS CCS}}
CCS processes are defined using simple operators (parallel composition, name prefixing or action, choice or sum, recursion, restriction and renaming), a single empty process denoted \(0\), and a collection of (co-)names \(a, \out{a}, b, \out{b}, \dots\) over which \(\lambda\) ranges:

\begin{equation}
	P, Q \coloneqq P\mid Q \BNFsepa \lambda . P \BNFsepa\sum_{i\in I} P_i %
	\BNFsepa A \BNFsepa P \bs a \BNFsepa P[a \leftarrow b] \BNFsepa 0 \tag{CCS Processes}
\end{equation}
where
\(A\) are (recursive) definitions of processes, that is \(A \overset{\text{def}}{=} P\) and \(A\) can occur in \(P\), \(P[a \leftarrow b]\) is the capture-avoiding substitution, and we write \(P[\namelist{a} \leftarrow \namelist{b}]\) for \(P[a_1 \leftarrow b_1, \hdots, a_n \leftarrow b_n]\).

To create SYS\textsubscript{CCS}, the processes are endowed with an evaluation mechanism given by the labeled transition system (LTS) of \autoref{fig:ltsrules_ccs}.
A process \(P\) reduces---or evaluates---to \(P'\) with label \(\alpha\) (which can be a (co-)name, or the special label \(\tau\) for \enquote{silent}, internal, transitions) if a tree whose root is \(P \redl{\alpha} P'\) can be derived using the rules of \autoref{fig:ltsrules_ccs}.

\begin{figure}
	{

		\centering

		\begin{prooftree}
			\hypo{P \redl{\alpha} P'}
			\infer1[com.\(_1\)]{P \mid Q \redl{\alpha} P' \mid Q}
		\end{prooftree}
		\hfill
		\begin{prooftree}
			\hypo{Q \redl{\alpha} Q'}
			\infer1[com.\(_2\)]{P \mid Q \redl{\alpha} P \mid Q'}
		\end{prooftree}
		\hfill
		\begin{prooftree}
			\hypo{P \redl{\lambda} P'}
			\hypo{Q \redl{\out{\lambda}} Q'}
			\infer2[syn.]{P \mid Q \redl{\tau} P' \mid Q'}
		\end{prooftree}
		\\[0.5em]
		\begin{prooftree}
			\hypo{}
			\infer1[act.]{\lambda . P \redl{\lambda}P}
		\end{prooftree}
		\hfill
		\begin{prooftree}
			\hypo{P_j \redl{\alpha} P_j'}
			\hypo{j\in I}
			\infer2[sum.]{\sum_{i\in I} P_i \redl{\alpha} P_i' }
		\end{prooftree}
		\hfill
		\begin{prooftree}
			\hypo{P \redl{\alpha} P'}
			\hypo{A \overset{\text{def}}{=} P}
			\infer2[rec.]{ A \redl{\alpha} P'}
		\end{prooftree}
		\\[0.5em]
		\begin{prooftree}
			\hypo{P \redl{\alpha} P'}
			\hypo{a \notin \alpha}
			\infer2[res.]{ P\bs a \redl{\alpha} P'\bs a}
		\end{prooftree}
		\qquad
		\begin{prooftree}
			\hypo{P \redl{\alpha} P'}
			\infer1[rel.]{P[\namelist{a} \leftarrow \namelist{b}] \redl{\alpha{[\namelist{a} \leftarrow \namelist{b}]} } P'[\namelist{a} \leftarrow \namelist{b}]}
		\end{prooftree}

	}
	\caption{Rules of the labeled transition system of CCS (LTS\textsubscript{CCS})}
	\label{fig:ltsrules_ccs}
\end{figure}

\subparagraph{Stage 2: \texorpdfstring{SPEC\textsubscript{CCS}}{SPEC CCS}}
The SYS\textsubscript{CCS} specification, although capturing the intended calculus, is cumbersome because of its syntactical rigidity: everything has to be spelled out rigorously, and basics properties like the commutativity of the product (i.e., writing \(P_1 \mid P_2\) or \(P_1 \mid P_2\) should not make a difference) are cumbersome to prove.
The solution is to define an equivalence relation on processes, define another specification using it---SPEC\textsubscript{CCS}---, and then to prove that they coincide.

\begin{definition}[CCS Structural equivalence]
	\label{def:equiv_ccs}
	Structural equivalence on processes is the smallest equivalence relation---sometimes \enquote{closed under context}---generated by the following rules:
	\begin{alignat*}{5}
		P \mid Q                    & \congru Q \mid P                                     & \qquad (P \mid Q)\mid V & \congru P \mid (Q \mid V) & \qquad P \mid 0 & \congru P %
		\\
		P + Q                       & \congru Q + P                                        & (P + Q) + V             & \congru P + ( Q + V)      & P + 0           & \congru P %
		\\
		(P \bs a) \mid Q            & \congru (P \mid Q) \bs a \text{ with }a\notin \fn{Q} & (P \bs a) \bs b         & \congru (P \bs b) \bs a                                 \\
		A \overset{\text{def}}{=} P & \Rightarrow A \congru P                              & P =_{\alpha} Q          & \Rightarrow P \congru Q
	\end{alignat*}
	Where \(\fn{Q}\) is the set of free names in \(Q\), and \( =_{\alpha}\) is the \(\alpha\)-equivalence given by capture-free substitution.
\end{definition}

SPEC\textsubscript{CCS} is then defined by adding the rule
\begin{prooftree}
	\hypo{P'_1 \congru P_1}
	\hypo{P_1 \redl{\alpha} P_2}
	\hypo{P_2 \congru P'_2}
	\infer3[con.]{P'_1 \redl{\alpha} P'_2}
\end{prooftree}
to LTS\textsubscript{CCS}, and removing com.\(_1\) (or com.\(_2\)), rec.\ and rel.\ from it.
Finally, the correctness of SPEC\textsubscript{CCS} is proven by showing that it is a \enquote{conservative extension of SYS\textsubscript{CCS}}, in the following sense:

\begin{lemma}
	\label{lem-ccs}
	If \(P\redl{\alpha} P'\) with SYS\textsubscript{CCS} and \(P\congru Q\) then \(Q\redl{\alpha} Q'\) with SPEC\textsubscript{CCS} and \(P'\congru Q'\).
\end{lemma}

This lemma guarantees that reasoning modulo the structural congruence does not alter the correctness of the system, and that we don't have to track small and innocuous modifications in the terms under study--it just cuts the red tape.

\section{What About Reversible Calculii?}

There are two ways of extending CCS to reversible calculus: the \enquote{static} approach \enquote{marks} each occurrence of an action, and gave birth to CCS with Communication Keys (CCSK)~\cite{Phillips2007b}.
The \enquote{dynamic} approach uses \emph{memories} attached to the threads of a process, and gave Reversible CCS (RCCS)~\cite{Danos2004,Danos2005}.
Both calculii evolved over the time, notably regarding structural equivalence.

\subparagraph*{CCSK}
Some versions of CCSK \enquote{does not exploit any structural congruence}\footnote{As noted in a paper~\cite{Lanese2019} that furthermore note that \(\alpha\)-equivalence can and should be added in any case~\cite[Remark 2]{Lanese2019}.}~\cite{Phillips2007b}, while some other~\cite{Graversen2018} uses a structural congruence that contains a significant subset of the one from CCS, plus a rule to deal with keys~\cite[p. 105]{Graversen2018}.

To our knowledge, the innocuity of this latter congruence has not been proven, but it was shown that a model based on reversible bundle structure was \enquote{stable}~\cite[Proposition 4.9]{Graversen2018} under it.

\subparagraph{RCCS}
RCCS' congruence evolved as well, from containing additive structural congruence but not the \(\alpha\)-equivalence~\cite[Section 2.3]{Danos2004} to being pretty much only the \(\alpha\)-equivalence~\cite{Aubert2016jlamp}.
But in any case, RCCS does not benefit from the clear distinction we discussed at the beginning of this paper. To understand why, let us explain how this system was designed.

In RCCS, reversibility is implemented thanks to memories that are \emph{complete}---so that every forward step can be backtracked---, \emph{minimal}--so that only the relevant information is saved---, but also \emph{distributed}---to avoid centralization.
Technically, \emph{memories} are attached to the threads of a process.
\begin{align}
	T    & \coloneqq m \rhd P \tag{Reversible Thread}                        \\
	R, S & \coloneqq T \BNFsepa R\mid S \BNFsepa R\bs a \tag{RCCS Processes}
\end{align}
for \(m\) a memory stack (whose definition~\cite{Aubert2016jlamp} is not included here), and \(P\) a CCS process.

A forward \emph{and} backward LTS is then defined, but it is extremely limited: any transition leading to \(m\rhd (P_1 \mid P_2)\) is blocked.
Indeed, to fulfill our requirement to distribute as much as possible the memories, the system needs to block this execution, and to be given an equivalence rule
\begin{equation}
	m\rhd (P_1 \mid P_2) \congru (\fork.m\rhd P_1) \mid (\fork.m\rhd P_2) \tag{Distribution of memory}\label{distrib-mem}
\end{equation}
that distributes the computation (and the memory, prepended with a special \enquote{fork} symbol \(\fork\)) over two units of computation.

There are two perspectives on this:
\begin{enumerate*}
	\item We can consider that the equivalence relation now becomes part of the definition of SYS\textsubscript{RCCS}, and that there are no SPEC\textsubscript{RCCS},
	\item We can consider that SPEC\textsubscript{RCCS} is strictly more expressive than SYS\textsubscript{RCCS}.
\end{enumerate*}

Both approaches are somehow problematic since we are losing a way of syntactically assessing that the equivalence relation (which generally contains more than just \ref{distrib-mem}) is \enquote{the right one}, and have to assume that it is reasonable.
The reduction semantics of CCS can also be considered to be SPEC\textsubscript{CCS}, so we loose the clear distinction between the two (reduction semantics and LTS) in RCCS.
Furthermore, to prove that RCCS and CCSK are equi-expressive~\cite{Lanese2019}, the authors had to remove %
from the rest of the LTS the possibility to use \ref{distrib-mem} between two transitions~\cite[Lemma 4]{Lanese2019}, making their proof actually \emph{harder}.
All in all, the questions we are asking are,
\begin{enumerate}
	\item Can we on one hand enrich this congruence (with \eg some form of commutativity of parallel composition), and on the other extract from it the \ref{distrib-mem} rule?
	\item What criteria shoud we use to determine that our newly defined SYS\textsubscript{RCCS} and SPEC\textsubscript{RCCS} are correct?
\end{enumerate}

Our first question aims at restoring the conceptual purity of our system and of the structural congruence, but also to provide tools to improve this latter (which is very limited, as of now).
Regarding our second question, and similarly to what has been done with CCSK, a model of the memories based on configuration structures has been show to be stable under structural congruence~\cite[Lemma 43]{Aubert2020a}.
However, the congruence considered was very small, and the result applied only to memories, and not to processes as a whole.

\section{Conclusion}

One of the true puzzle raised by this small introspection is the second question asked previously: while \enquote{making up} SYS\textsubscript{RCCS} and SPEC\textsubscript{RCCS} is \enquote{easy}, gaining evidence that they are the correct systems, and that they are equi-expressive, is harder.
The equivalence between CCSK and RCCS, as well as the numerous denotational models for those calculi and their simulation results, are good safe-guards that those models are relevant, and capture reversibility in a satisfactory way.
But how can we endow them with \enquote{the right} structural congruence?

This investigation actually raises another question, less expected: looking back, the way CCS asses that its equivalence relation is \enquote{the right one} is somehow circular, since our \autoref{lem-ccs} uses the equivalence \enquote{on the outside} \emph{and} \enquote{on the inside}.
One of our current investigation is to precisely pinpoint why is this equivalence in CCS acceptable, and to see if the criteria we identify can be imported to RCCS and CCSK.
Mechanized methods are likely to be of little help in this domain\footnote{An impressive list of the \enquote{mechanized formalizations of the \(\pi\)-calculus (and friends)} can be found at \url{http://why-lambda.blogspot.com/2015/04/mechanized-formalizations-of-pi.html}.} as \eg formalization of CCS in Coq~\cite{Affeldt2008} or of the \(\pi\)-calculus in Agda~\cite{Perera2018} either discard syntactical congruences, or pose them as axioms.
\bibliographystyle{eptcs}
\bibliography{standalone}
\end{document}